# Investigation of Edge States in Artificial Graphene Nano-Flakes

Qiushi Zhang[1,2], Tszchun Wu[1], Guowen Kuang[1], A'yu Xie[1], and Nian Lin[1]*

[1]Department of Physics, The Hong Kong University of Science and Technology, Hong Kong, China

[2]Department of Aerospace and Mechanical Engineering, University of Notre Dame, Notre Dame, IN, USA

*Correspondence to: phnlin@ust.hk

**ABSTRACT**

Graphene nano-flakes (GNFs) are predicted to host spin-polarized metallic edge states, which are envisioned for exploration of spintronics at the nanometer scale. To date, experimental realization of GNFs is only in its infancy because of the limitation of precise cutting or synthesizing methods at the nanometer scale. Here, we use low temperature scanning tunneling microscope (STM) to manipulate coronene molecules on a Cu(111) surface to build artificial triangular and hexagonal GNFs with either zigzag or armchair type of edges. We observe that the metallic edge states only exist in the GNFs with zigzag edge and localize at the most outside one type of the sublattice. The experimental results agree well with the tight-binding calculations. To our knowledge, our work renders the first systematic experimental confirmation of the predicated electronic properties of the GNFs.

**INTRODUCTION**

Graphene has attracted a lot of scientific interests in the past few decades[1–4]. The low-energy quasi-particles can be described as massless Dirac fermions with the two pseudospin components contributed by the two different types of sublattices[5–8]. Following the successful experimental fabrication, rich electronic and chemical properties of graphene-based materials were uncovered[9–13]. When cutting a piece of graphene sheet into nanometer size, many new properties may emerge[14–17]. Particularly, the so-called graphene nanoribbons (GNRs) and graphene nano-flakes (GNFs) are featured with remarkable electronic properties such as spin-polarized metallic edge states and band gap[14,18–28]. GNFs are anticipated to exhibit a richer variety of electronic properties than GNRs due to the unique features in their shape, size and edge morphology. For instance, the quantum dots mimicked by GNFs are featured with extremely long spin relaxation and decoherence time due to small spin-orbit and hyperfine coupling in carbon atom[29–33]. Additionally, as a result of the specific boundary conditions involved in the Dirac equations, the energy spectrum of the GNFs can be non-trivially tuned by edge type[14–17,30,31]. Generally speaking, there are two fundamental types of edges in GNFs, the zigzag and arm-chair ones. The GNFs with zigzag edges exhibit a half-filled flat band at the Fermi level, which renders a zero-energy state locating at the edges of GNFs. Based on theoretical works[30,31,36–41], the electronic properties, such as the density of states (DOS) and the metallic edge states, differ in GNFs with distinctive shapes, sizes and types of edges[34,35,42–44].

Compared with numerous theoretical works, the experimental progresses in fabricating and characterizing GNFs are relatively slow. Although some GNFs samples have been successfully manufactured by the soft-landing mass spectrometry[45], it is still very challenging to accurately control the shape, size or edge type with reasonable repetition rate[46,47]. The main obstacle is to cut out or synthesize GNFs with precise edge morphology at the nanometer scale. An alternative approach to examine the physics of GNFs is to fabricate artificial graphene systems that are made out of ultra-cold atom[48,49], nanopatterned quantum wells[50,51] or artificial molecular system[52–54]. The later one, artificial molecular system was firstly proposed by C.-H. Park and S. G. Louie in 2009[55]. By STM tip manipulating, a feasible method widely used in on-surface molecular study[56–58], CO[59–61] or coronene molecules[52,53] are patterned on a Cu(111) surface to create potential lattice, the two-dimensional electron gases (2DEGs) of metal surface are confined and scattered. A triangular nano-potential lattice results in the 2D massless Dirac system which is considered to behave equivalently to those of the low-energy charge carriers in real graphene layer[55,61–64]. For instance, K. K. Gomes et al. successfully demonstrate the quantum tunneling effect between atomically sharp junctions and Landau levels in pseudo magnetic fields[59] with the artificial graphene systems built by CO atoms. In this work, we build artificial GNFs (AGNFs) with different geometries aiming to study the metallic edge states in the GNFs. Specifically, we construct the triangular and hexagonal AGNFs with either zigzag or armchair edges. We confirm that the edge state can only be observed in the AGNFs with the zigzag edges and the edge states are localized at the outmost sublattice sites.

**RESULTS AND DISCUSSION**

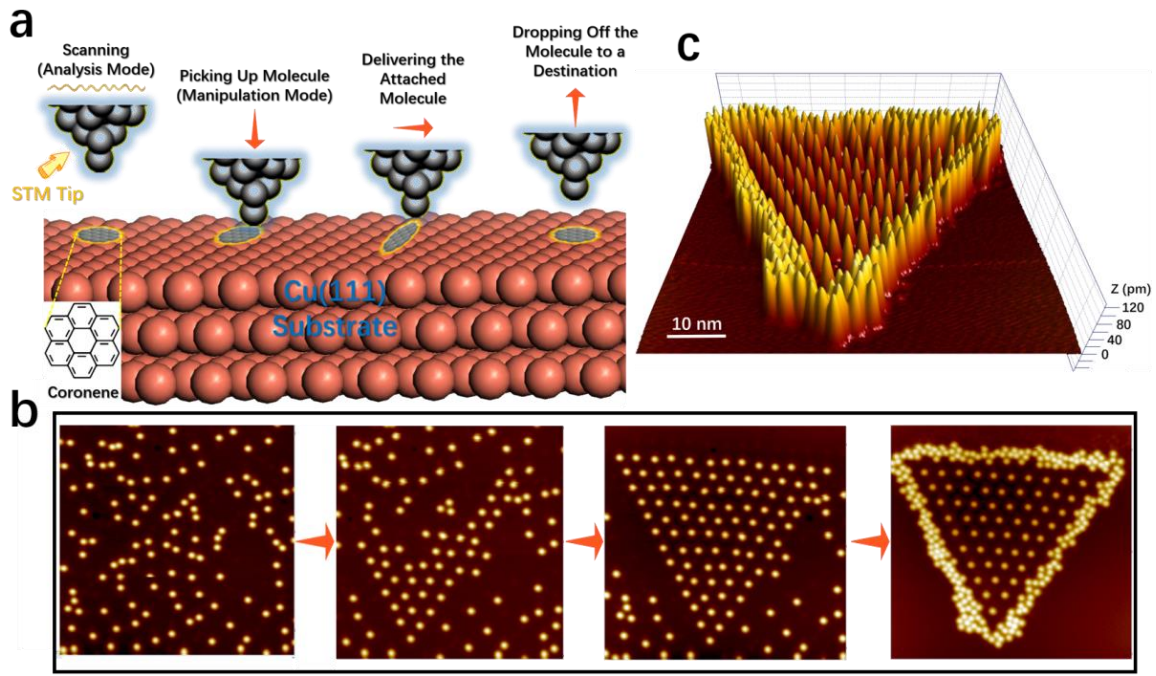

**Figure 1.** (a) Schematic of STM manipulation of coronene molecules on a Cu(111) surface. (b) STM images presenting the construction process of a triangular AGNF with zigzag edges (Setpoint: -1 V, 0.3 nA, 46×46 nm$^2$). (c) The 3D STM image of the last panel in (b).

We first deposit coronene molecules on a Cu(111) substrate. The molecules are randomly distributed on the surface (see Methods section for details). **Figure 1a** demonstrates the STM manipulation process carried out at 4.8 K: The STM tip is moved down from the normal scanning height to approach an adsorbed coronene molecule. The distance between the tip and coronene is within ~2 Å at this moment. When a bias voltage of -2 V is applied between the STM tip and the Cu(111) substrate, the coronene molecule is lifted up to the tip. Then, the lifted molecule is laterally dragged to a desired location. The tip is set back to the normal scanning height and the molecule is dropped off from the tip onto the destination on substrate. By repeating this manipulation process, we can move the randomly distributed coronene molecules one by one into a designed shape. **Figure 1b** presents the step-by-step construction of a triangular AGNF with zigzag edges. Finally, a molecular potential wall consisting of many molecules is built surrounding the AGNF to confine the 2DEGs, forming an AGNF, as shown in **Figure 1c**.

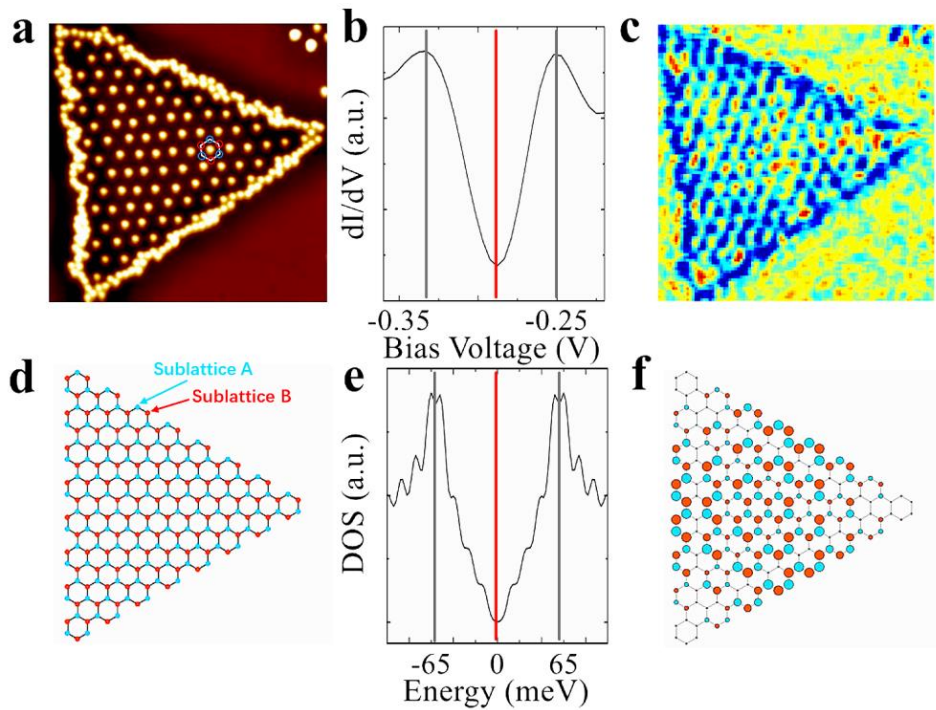

**Figure 2.** (a) STM image of a triangular AGNF with armchair edges (Set-point: -1 V, 0.3 nA, 46×46 nm$^2$). Inset: Blue (red) circles denote sublattice-A (B) atoms, and the hexagonal frame represents a six-member ring of graphene. (b) dI/dV spectrum averaged over the whole AGNF area excluding the coronene sites and subtracting substrate background (Set-point: -1 V, 0.3 nA). The red line indicates the Dirac point energy. (c) STS map (46×46 nm$^2$) acquired at the Dirac point (-0.29 V). (d) Atomic structure of the equivalent GNF. The sublattice A (B) atoms are colored in blue (red). (e) and (f) Tight-binding (TB) calculation results of the structure shown in (d): total DOS (e), spatial distribution of DOS at the Dirac point (E = 0 eV) (f). The size of the dots is proportional to the DOS intensity in (f).

**Figure 2a** shows an equal-lateral triangular AGNF with armchair edges. The lattice constant is ~3.0 nm, and the length of each side of the triangular flake is ~40 nm. The color circles overlaid in **Figure 2a** highlight the artificial "carbon atoms", in which blue (red) circles denote sublattice-A (B) atoms, and the hexagonal frame represents a six-member ring of a graphene layer. The atomic model of an equivalent GNF is presented in **Figure 2d**, in which the two sublattices, A and B, are colored in blue and red, respectively. **Figure 2b** shows a dI/dV spectrum averaged over the whole area of this AGNF, excluding the coronene molecules and subtracting the substrate background. The spectrum displays a V shape sided by two peaks at -0.35 and -0.25 V, which signifies the DOS of the massless Dirac fermions. The valley of the V-shape, which corresponds to the Dirac point[55,55,62,63], locates at -0.29 V[52]. The spatial distribution of the DOS at the Dirac point (-0.29 V) is plotted in **Figure 2c**. The density spreads over the whole AGNF without any edge-state feature. Theoretically, we use a tight-binding (TB) model to calculate the electronic properties of the structure shown in **Figure 2d**. Due to the electron-hole symmetry, the energy level of the Dirac ponit is at 0 eV. The calculated DOS shown in **Figure 2e** shows a symmetric V shape with two peaks located ~60 meV below and above the Dirac point. The calculated DOS nicely agrees with the dI/dV spectrum in **Figure 2b**. **Figure 2f** shows the spatial distribution of DOS at the Dirac point, showing that the DOS spreads over the whole GNF, similar as the experimental result shown in **Figure 2c**.

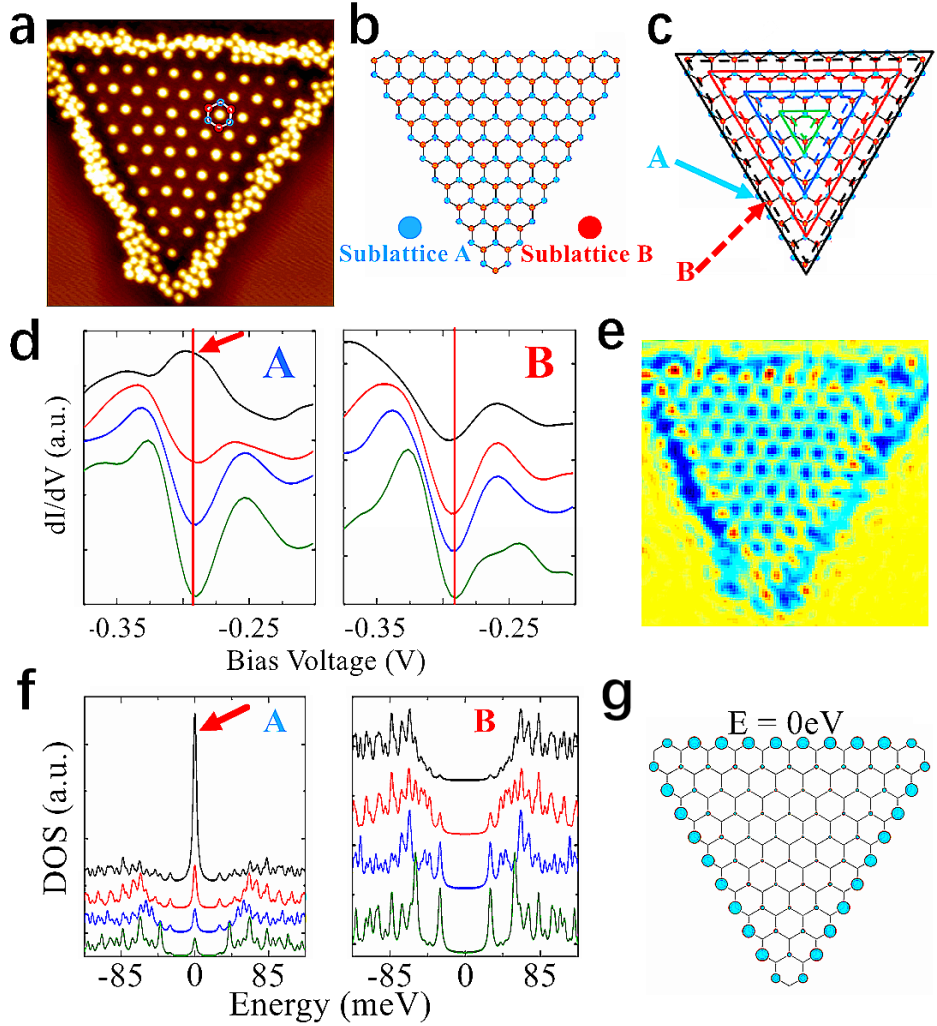

**Figure 3.** (a) STM image (40×40 nm$^2$) of a triangular AGNF with zigzag edges (Set-point: -1 V, 0.3 nA). Inset: Blue (red) circles denote sublattice-A (B) atoms, and the hexagonal frame represents a six-member ring of graphene. (b) Atomic structure of the equivalent GNF. The sublattice A (B) atoms are colored in blue (red). (c) Schematic shows four layers of equi-lateral triangles in this GNF from edges to center. The solid (dash) lines mark the atoms of sublattices A (B). (d) Site-specific dI/dV acquired at the different triangle layers in the AGNF. The color code follows the definition given in (c). (e) STS map (40×40 nm$^2$) acquired at the Dirac point (−0.29 V). (f) and (g) TB calculation results of the structure shown in (b): (f) Site-specific local DOS of the different triangle layers in the GNF. The color code follows the definition given in (c). (g) Spatial distribution of DOS at the Dirac point (E = 0 eV). The size of the dots is proportional to the DOS intensity.

**Figure 3a** shows a triangular AGNF with zigzag edges. The lattice constant and flake size are same as those of the armchair AGNF shown in **Figure 2a**. The atomic model of an equivalent GNF is presented in **Figure 3b.** The two sublattices are inequivalent in a zigzag edge triangular GNF. This is illustrated in **Figure 3c**, where the atoms of the sublattices A and B sit on the solid and dashed triangles, respectively. We define the atoms sitting at the same triangles as equivalent site atoms. The equivalent site averaged dI/dV spectra of the different triangle layers are plotted in **Figure 3d.** The colors follow the same definition as in **Figure 3c:** The black, red, blue and green curves in the right (left) panel are the spectra of the A (B) sublattice atoms sitting at the triangles of the same color in **Figure 3c**. **Figure 3d** shows: the sublattice-A sites at the outmost triangle feature a peak at the Dirac point of -0.29 V, and all other sites, including the inner triangles of the sublattice-A sites and all the sublattice-B sites, feature a V-shape valley at the Dirac point. **Figure 3e** shows the state at -0.29 V is localized at the three zigzag edges of the AGNF. We then perform TB calculation on a triangular zigzag-edge GNF of the same size. **Figure 3f** shows the site-specific local DOS, revealing that the sublattice A atoms feature a sharp peak at the Dirac point and the peak intensity decays from the edges to the interior rapidly. In contrast, this peak is absent in the sublattice B atoms. **Figure 3g** displays the spatial distribution of the DOS at the Dirac ponit, showing that only sublattice-A atoms exhibit appreciable DOS, and the three edges feature highest DOS intensity. Overall, the TB calculation captures the major characteristics of the edge state observed in the experiments.

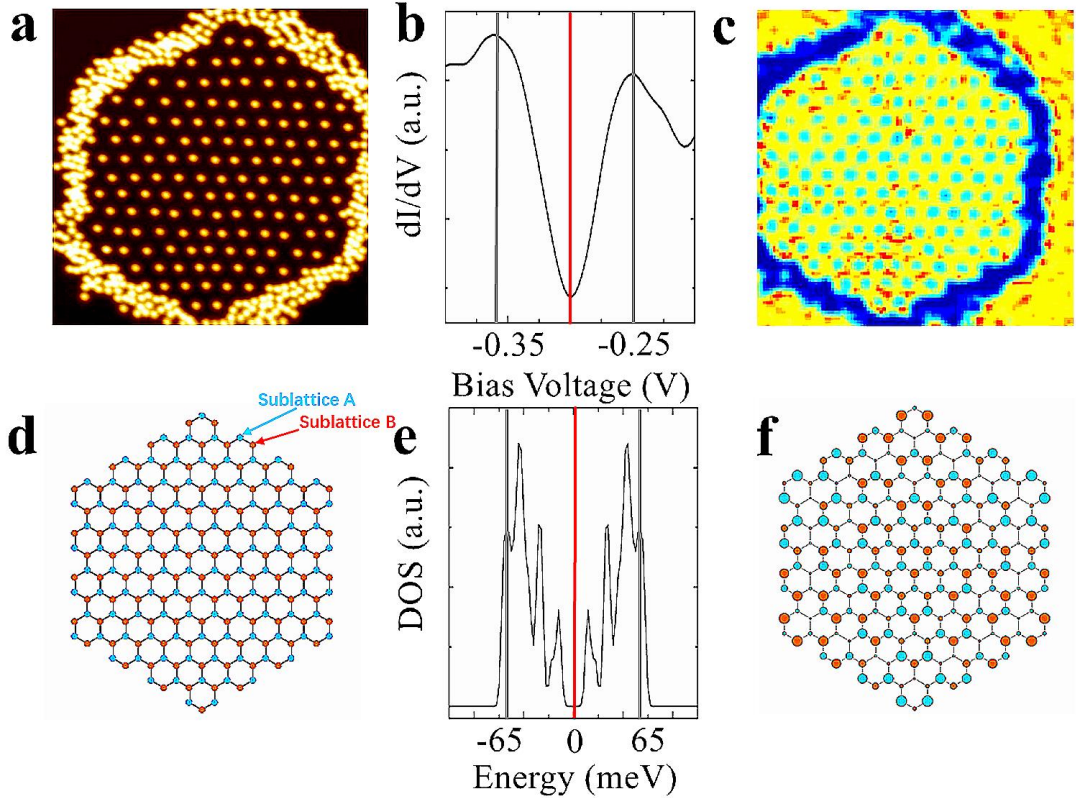

**Figure 4.** (a) STM image (48×48 nm$^2$) of a hexagonal AGNF with armchair edges (Set-point: -1 V, 0.3 nA). (b) dI/dV spectrum averaged over the whole AGNF area excluding the coronene sites and subtracting substrate background (Set-point: -1 V, 0.3 nA). The red line indicates the Dirac point energy. (c) STS map (46×46 nm$^2$) acquired at the Dirac point (-0.30 V). (d) Atomic structure of the equivalent GNF. The sublattice A (B) atoms are colored in blue (red). (e) and (f) TB calculation results of the structure shown in (d): total DOS (e), spatial distribution of DOS at the Dirac point (E = 0 eV) (f). The size of the dots is proportional to the DOS intensity.

**Figure 4a** shows a hexagonal armchair-edge AGNF. Similar to the triangular armchair-edge AGNF, the averaged dI/dV spectrum is a V shape, as shown in **Figure 4b**, where the Dirac point is located at -0.3 V. **Figure 4c** displays spatial distribution of DOS at the Dirac point, showing the density spreads over the entire AGNF without any edge-state feature. These results corroborate the TB calculation of the model shown in **Figure 4d**. The calculated total DOS presented in **Figure 4e** nicely reproduces the V shape, whereas the spatial distribution of DOS at E = 0 eV plotted in **Figure 4f** reveals a uniformly distributed DOS.

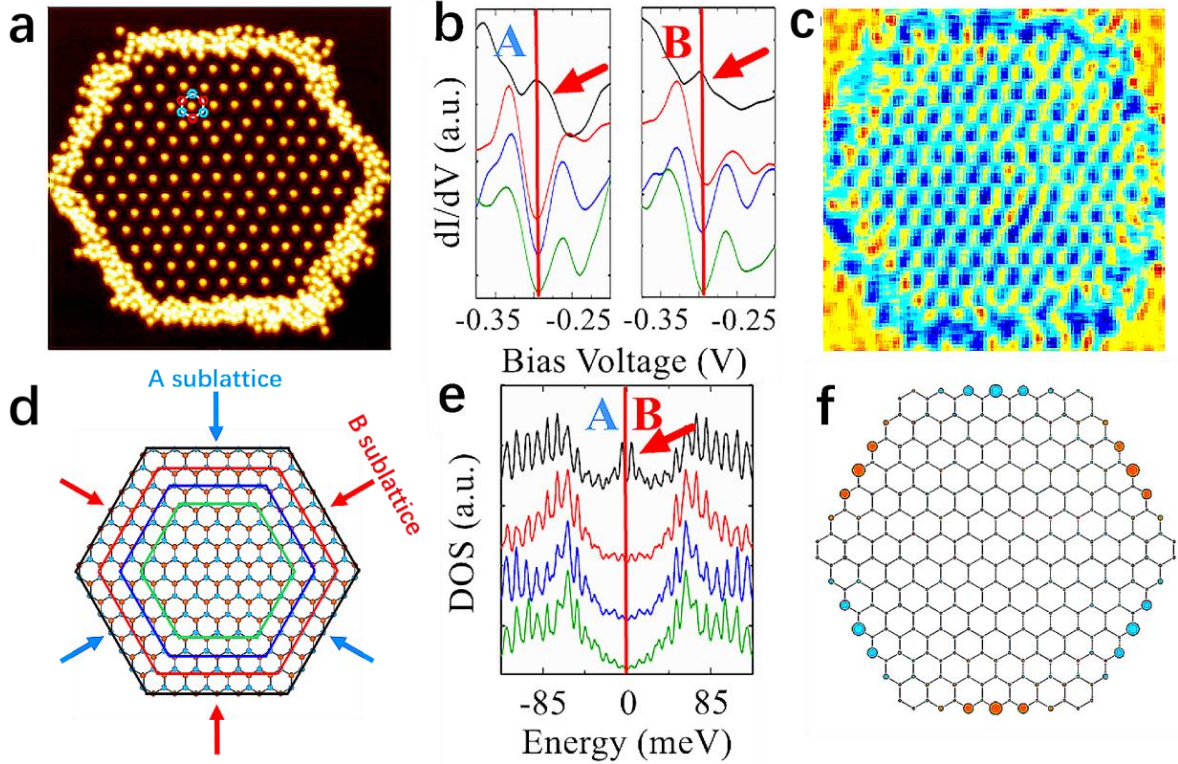

**Figure 5.** (a) STM image (50×50 nm$^2$) of a hexagonal AGNF with zigzag edges (Set-point: -1 V, 0.3 nA). (b) Site-specific dI/dV acquired at the sites sitting at the different hexgons in the AGNF. The color code follows the definition given in (d). (c) STS map (50×50 nm$^2$) acquired at the Dirac point (−0.29 V). (d) Atomic structure of the equivalent GNF. The sublattice A (B) atoms are colored in blue (red). Four layers of hexgons define the equivalent sites. (e) and (f) TB calculation results of the structure shown in (d): (e) Site-specific local DOS of the atoms at the different hexgons in the GNF. The color code follows the definition given in (d). (f) Spatial distribution of DOS at the Dirac point (E = 0 eV). The size of the dots is proportional to the DOS intensity.

Last, we present the results of a hexagonal zigzag-edge AGNF as shown in **Figure 5a**. The atomic model of an equivalent GNF is presented in **Figure 5d**, illustrating three edges are A sublattice while the other three edges are equivalently B sublattice. The colored hexgons define the equivalent site atoms of the two sublattices. **Figure 5b** shows the dI/dV spectra acquired at the sites sitting at different hexgons while the colors follow the same definition as in **Figure 5d:** The black, red, blue and green curves in the right (left) panel are the spectra of the A (B) sublattice atoms sitting at the hexagons of the same colors in **Figure 5d**. The sites at the outmost hexgon of the both sublattices feature a peak at the Dirac point and the sites at all the inner hexgons feature a V-shape valley at the Dirac point. **Figure 5c** shows the DOS at -0.29 V is localized at the six zigzag edges of the AGNF. We perform TB calculation on the structure shown in **Figure 5d.** Both sublattices exhibit same site-specific local DOS as shown in **Figure 5e**: A peak at the Dirac point is presented only at the outmost hexgon. **Figure 5f** reveals that the DOS at the Dirac point (E = 0 eV) are mainly confined at the six edges, identical for the two sublattices. This symmetric behavior is given by the symmetric geometry of the two sublattices in the hexagonal GNFs.

## CONCLUSIONS

In conclusion, we present an experimental study to evidence the metallic edge states in the AGNFs. The comparisons between the zigzag-edge and armchair-edge AGNFs of triangular and hexagonal shapes confirm that the edge states at the Dirac point only exist at the outmost type of sublattice at the zigzag edges.

## EXPERIMENTAL AND THEORETICAL METHODS

**Sample Preparation:** The experiments were performed in an ultrahigh vacuum scanning tunneling microscope (STM) integrated with scanning tunneling spectroscopy (STS) systems (Scienta Omicron) with a base pressure of $8 \times 10^{-10}$ mbar. Liquid helium is used to keep the operation temperature constantly at 4.8 K. The Cu(111) single-crystalline substrate was cleaned by several cycles of $Ar^+$ sputtering and annealing. Coronene molecules (Sigma Aldrich) were thermally evaporated onto the Cu(111) substrate at the room temperature by using an organic molecular evaporator. STS mapping was performed to acquire spatially resolved density of states (DOS) of the whole AGNFs. The STS spectra were measured using a lock-in amplifier with a sine modulation of 1.5 kHz and a modulation of 4 mV. Each dI/dV spectrum was normalized by (I/V) data to obtain the DOS or site-specific local DOS plot. All STM and STS data were acquired in a constant current mode. A bias voltage of -2 V is applied between the STM tip and Cu(111) substrate, which enables the tip to lift up a single coronene molecule. Additional coronene molecules were packed closely to form walls surrounding the lattice to construct

AGNFs. The orientation of the walls with respect to the lattice defines the edge morphology of the AGNFs, i.e. zigzag or arm-chair edges.

**TB Calculation:** We also carried out the tight-binding (TB) calculations with a nearest-neighboring hopping parameter of 65 meV using the Python package (pybinding). The hopping parameter ($t$) is obtained with the formula: $t = v_F \frac{2\hbar}{\sqrt{3}d}$, where $d$ is the lattice constant (3 nm), $\hbar$ is the reduced Planck constant, and $v_F$ is the Fermi velocity ($2.6 \times 10^5$ m/s, the Fermi velocity of AGNFs system is calculated by the formula: $v_F = \frac{2\pi\hbar}{3m^*d}$, where $m$ is the effective mass of the Cu surface-state band is 0.4 $m_e$[52]). The site-specific local DOS is obtained through the Green function, which is derived from the Kernel polynomial method with a Gaussian broadening of 2.3 meV. The DOS is calculated from the eigenvalues with the same Gaussian broadening.

## ACKNOWLEDGEMENTS

This work is supported by Hong Kong RGC (No. 16304016).

## AUTHOR CONTRIBUTIONS

Q. Z., G. K. and N. L. designed the experiments, and Q. Z. and G. K. set up the experiments. Q. Z. and G. K. performed the experiment. Q. Z., T. W. and A. X. designed the simulations and T. W., and A. X. performed the simulations. Q. Z. and N. L. wrote the manuscript, T. W., G. K., and A. X. revised it.

## COMPETING INTERESTS

The authors declare no conflict of interest.